# Magneto-optical extinction trend inversion in ferrofluids

S. I. Shulyma, B. M. Tanygin, V. F. Kovalenko, M. V. Petrychuk

*Faculty of Radiophysics, Taras Shevchenko National University of Kyiv, 4G, Acad. Glushkov Ave., Kyiv, Ukraine,*

*03187*

ABSTRACT

Effects of pulse magnetic field on the optical transmission properties of thin ferrofluid (FF) layers were experimentally investigated. It was observed that, under an influence of an external uniform magnetic field, pulses applied to the samples surfaces in normal direction decrease the optical transmission with further returning it to its original state, even before the end of the field pulse. The dependencies of the observed effects on the magnetic pulse magnitude and the samples thickness were investigated. The experimental results are explained using FF columnar aggregates growth and lateral coalescence under influence of a magnetic field, leading to a light scattering type Rayleigh-to-Mie transition. Further evolution of this process comes to a geometrical optics scale and respective macroscopic observable opaque FF columnar aggregates emergence. These changes of optical transmission are non-monotonic during the magnetic field pulse duration with minimal value in the case of Mie scattering, which is known as a magneto-optical extinction trend inversion. The residual inversion was detected after the external magnetic field pulse falling edge. Using molecular dynamics simulation, we showed that a homogeneous external magnetic field is enough for the formation of columnar aggregates and their fusion. The results clarify the known Li theory (Li et al., J. Phys. D: Appl. Phys. 37 (2004) 3357, and Sci. Technol. Adv. Mate. 8 (2007) 448), implying an inhomogeneous field as a required prerequisite for the magneto-optical extinction trend inversion phenomenon.

*Email addresses:* kiw_88@mail.ru (S. I. Shulyma), b.m.tanygin@gmail.com (B. M. Tanygin)

**Introduction**

Ferrofluids (FFs) are colloidal suspensions of magnetic nanoparticles in a carrier liquid. Increasing interest to FFs is related to their important applications, for instance in drug deliveries [1], hyperthermia [2], magneto-resonance tomography [3], and some others [4,5]. Additionally, a method for fundamental understanding of the phase transitions can be developed using FF as a model system. The FF aggregates can emerge even in the absence of a magnetic field [6,7]. An emergence of the FF aggregates [8-12] predictably leads to changes in its optical and other physical properties [13–15].

Even a single ferromagnetic nanoparticle reveals complex light scattering and absorption properties [16]. An FF magneto-optical extinction [17] has been investigated with considerations of aggregation phenomenon in the sample subjected to a normal magnetic field [18–22]. The reversible zippering of nanoparticle chains in magnetic FFs under external magnetic fields and corresponding time-dependent transmitted light intensity and the scattered pattern were studied [23]. It was experimentally shown that the magneto-optical extinction trend inversion (ETI) is caused by formation of drop-like aggregates [19]. When a laser with a wavelength of $\lambda$, much larger than a nanoparticle diameter, is used, then an FF aggregate can be modeled using a continuous medium with a classical electrodynamics including scattering theories.

It was shown that light illumination of thin FF layers ($\sim$ 200 $\mu$m) induces progressive isotropic formation of magnetic aggregates [15]. The planar weak magnetic field ($\sim$ 10 Oe) induces the anisotropic growth of the aggregates for $\sim$ 60 s, which changes the FF optical transmission. It was shown that short planar magnetic field pulses ($\sim$ 5 s) applied to thick FF layers ($\sim$ 10 mm) with aggregates induces a sharp increase in the scattered light intensity right after the pulse rising edge [24]. This value did not change over the pulse duration and returned back to the initial magnitude after the pulse falling edge. The characteristic time of the scattered light signal rise was observed larger than one of the signal fall by two orders of magnitude.

Hence, an optical radiation can be used as a detector of an emergence and change of an FF magnetic aggregates structure. These results are important for basic understanding of the aggregates formation and investigation on the FF functional properties. However, the literature review indicates that so far no simulation-based investigation have been performed on the role of FF aggregates internal structural changes in the magneto-optical extinction experimental phenomena. We used a molecular dynamics simulation (nanoparticle Brownian motion level) method in this study [25–28]. The present work is focused on the computational and experimental research.

## 1. Materials and methods

### 1.1. Experimental setup

The $Fe_3O_4$ FF sample was prepared using precipitation method [29,30]. Aqueous solutions of iron salts were prepared through dissolution 27 g $FeCl_3 \cdot 6H_2O$ and 15 g $FeSO_4 \cdot 7H_2O$ (Makrohim, Ukraine) each in 250 mL of deionized water and were separately filtered at room temperature. Afterwards, they were intensively mixed and gradually over a period of 5 min 125 mL of 25% ammonium hydroxide solution in water (Makrohim, Ukraine) was added and then diluted with 250 mL of deionized water. As a result of chemical reactions, a black precipitate was formed. To remove the water-soluble salts from the product, it was washed 5 times with deionized water. Then, the magnetite water suspension was heated to 75 °C and supplemented with 15 mL oleic acid (Makrohim, Ukraine) and 75 mL kerosene TS-1 (aviation fuel). Next, the vessel with the sample was rested for 2 h, resulting in a black precipitate with a top liquid phase that was decanted away. The remaining black precipitate was washed 4 times with deionized water and then 2.5 mL oleic acid and 25 mL kerosene were added. The leftover water was evaporated on a bain-marie. For optical scattering measurements, the FF sample was diluted in deionized water to a volume fraction of $\varphi_V \sim 1.2\%$.

The investigated FF possessed a high sedimentation stability; an aggregation was observed in an optical microscope only when external magnetic field was applied. The magnetic field switch-off led to destruction of aggregates and transformation of FF system into a conventional uniform FF suspension consisting of separate ferromagnetic nanoparticles or small invisible in optical microscope aggregates. Experimental procedure consisted of FF samples entrapped in the space formed between a microscope slide and a cover slip as a gap. The space thickness $h$ was controlled and restricted using copper wires with known diameters of 50, 120, and 250 $\mu$m.

The optical transmission experimental setup is outlined in Fig. 1. A system of He–Ne laser LGN-208A (Polyaron, Lviv) was used as a source of optical radiation with a beam diameter of $\sim 1$ mm. The electromagnet (#2) was inducing a magnetic field along the optical radiation direction of Z. The FF sample (#3) was placed in the gap between electromagnet poles and the microscope slide (plane XY) oriented orthogonally to magnetic field force lines and incident laser beam. The amount of field spatial inhomogeneity in the sample vicinity and laser beam area were less than 4% and 1% respectively. The electromagnetic winding used for varying the electrical current value was allowed to change the applied magnetic field magnitude $H_0$ in the range of 0–3 kOe measured with a solid-state 4 mm × 3 mm Hall-effect sensor SS490 (Honeywell, USA). The magnetic field pulse was produced using an electromechanical switch.

The laser beam was directed to the beam splitter (#4), allowing the simultaneous transmission coefficient measurement and the FF layer visual observation with a photodetector (#5) and CMOS USB camera (#6), respectively. The camera was attached to an optical microscope. This optical system magnification range was 10×..400×. Here, only FF structure with granularity size $>\lambda/2$ could be recognized. The photodetector electric signal was transmitted to a personal computer (PC) using a voltage registration with an external analog-to-digital converter Triton 3000 U (Terex, Ukraine) with time samples divisions of 150 ms.

Experimental data provided in the present manuscript have been measured at first pulse cycle. However, additionally, optical transmission value has been monitored for several additional cycles. A supposed return of FF to same initial state was achieved by introducing a time interval between measurements (5 minutes). The sonication technique has not been used. Here, after any measurement the FF structure was monitored using the optical microscope and using laser light scattering parameters.

Simulation of a polydisperse FF requires determination of nanoparticle radius $r$ probability density function $f(r)$. Two analytical techniques were used for this purpose, transmission electron microscopy (FESEM, MIRA3 TESCAN, USA) in TEM mode (operating voltage 20 kV) and dynamic light scattering (Zeta Sizer ZS, Malvern Instruments, UK). The TEM experimental sample was prepared through putting an FF drop on a TEM grid support film of Carbon/Formvar/Copper type. The kerosene solvent was evaporated in vacuum at room temperature.

*1.2. Simulation*

A molecular dynamics method for simulation [7,27,28] was used. All simulation runs were repeated 4 times with random initial conditions. The results were averaged. A Cartesian coordinate system of XYZ coinciding with the one defined in the experimental setup section was used. A real thickness of the experimental sample being investigated here corresponds to large number ($N$) of FF nanoparticles leading finally to durable simulation of an FF aggregate formation evolution. The present method corresponds to machine calculation timeframes $\sim N^2$. In order to optimize a calculation time, the spatial cell $0.32 \times 0.95 \times 2.85 \mu$m and $N = 750$ have been used in the simulation, which is much less than a real experimental sample parameters. Hence, corresponding approximate translational symmetry is supposed. Possible restriction of such model is a real sample surface demagnetization field effects. However, as far as the FF sample was diluted by deionized water to the low volume fraction ($\varphi_V \sim 1.2\%$), according to estimations, an effect of a demagnetization field (bulk plate geometry) can be neglected. The periodic boundary conditions with minimum-image convention have been disabled in order to simulate a compressive nature of the columnar aggregates creation: nanoparticles concentrate into a geometrical center of the experimental sample even in a homogeneous external magnetic field. Also, aggregates mutual interaction is short-range and weaker than a dipole interaction. Hence, periodic boundary conditions do not impact overall process significantly.

The translational and rotational motion finite-difference Euler scheme is based on the original method [7,27,28]. At some fixed timestamp, each modeled nanoparticle is considered as a single domain ferromagnetic nanoparticle with a magnetic moment $m_i$, where $|m_i| = const$. The simulation method directly takes into account a relation of the selected molecular dynamics simulation time step $\Delta t = 1.5 \cdot 10^{-8}$ s with following temporal values:

1. The Néel relaxation theory attempt time $\tau_0 = 10^{-10} - 10^{-9}$ s

2. The Néel relaxation time $\tau_N$ as a function of the nanoparticle diameter $d$

3. Brownian relaxation time $\tau_B$ as a function of the nanoparticle diameter $d$

The dependencies $\tau_N(d)$ and $\tau_B(d)$ have been taken from the reference theoretical and experimental data [31]. Depending on these relations, each nanoparticle was modelled in a special way. Precise description of this algorithm is provided in the [27]. In case of relation $\tau_0 < \tau_N < \Delta t$ the magnetic moment was aligned with the total magnetic field direction in the nanoparticle center at each simulation step (Néel relaxation flip). In case of relation $\tau_0 < \Delta t < \tau_B < \tau_N$ the rotational motion of the nanoparticle has been simulated by the finite-difference Euler scheme taking into account internal nanoparticle magnetocrystalline anisotropy $-KV^m \cos^2 \theta$, where $\theta$ is an angle between the easy magnetization axis and the magnetic moment $\mathbf{m}$; $K$ is a nanoparticle magnetic anisotropy constant [32]; and $V^m$ is a volume of the nanoparticle ferromagnetic core. Here, the magnetic moment $\mathbf{m}$ was instantly aligned with the effective field due to relation $\tau_0 \ll \Delta t$, meaning that an internal structure of the nanoparticle is equilibrium [32].

In scope of the continuum solvation approximation of not very dense solutions, the Brownian translational and rotational nanoparticle motion is described by the Langevin equations with the hydrodynamic-originated Langevin parameters [27,33]:

$$M \mathrm{d}^2 \mathbf{r}/\mathrm{d}t^2 = \mathbf{F} + \mathbf{R}^T \qquad (1)$$

$$I \mathrm{d}\boldsymbol{\omega}/\mathrm{d}t = \mathbf{T} + \mathbf{R}^R \qquad (2)$$

where $M$ and $I$ are the nanoparticle mass and moment of inertia respectively; $\mathbf{F}$ and $\mathbf{T}$ are the total effective force and the torque acting on the given nanoparticle; $\boldsymbol{\omega}$ is a nanoparticle self-rotation angular velocity; $\mathbf{R^T}$ and $\mathbf{R^R}$ are the random force and torque respectively, which have been modelized by Gaussian noise:

$$<\mathbf{R^T}(t)> = 0 \qquad (3)$$

$$\langle \mathbf{R^T}(t)\mathbf{R^T}(t')\rangle = 6kT\gamma^T\delta(t-t') \tag{4}$$

$$\langle \mathbf{R^R}(t)\rangle = 0 \tag{5}$$

$$\langle \mathbf{R^R}(t)\mathbf{R^R}(t')\rangle = 6kT\gamma^R\delta(t-t'), \tag{6}$$

where damping coefficients of translational and rotational motion are given by $\gamma^T = 3\pi\eta d$ and $\gamma^R = 6V\eta$, where $\eta$ is a carrier liquid dynamic viscosity and $V$ is a total hydrodynamic nanoparticle volume. Equations (1) and (2) has been integrated in the scope of the viscous limit approximation [7,27,34].

The uniform FF suspension corresponds to the nanoparticle coating by oleic acid molecules $k_c = N_S/N_S^{max} \approx 50\%$, where $N_S$ is a surface density of molecules number on the nanoparticle and $N_S^{max} = 2 \cdot 10^{18}$ m$^{-2}$ is its maximal value for the case of the oleic acid molecules and a magnetite nanoparticle [31]. Only the largest nanoparticles which form aggregates (including nanoparticles linear chains) were considered [7,27]. The mole fractions of these nanoparticles were phenomenologically observed in the range of $\Phi_L \sim 5\%-10\%$ [35]. The largest nanoparticles decisively determine the overall phase transition [35]. It can be assumed that the other nanoparticles weakly affect the formation of structures. In the present simulation, in order to optimize a computational time, only the largest nanoparticles were modeled. This assumption should be validated in a scope of further all-particles scale simulations through the Graphics Processing Units based optimization method [36,37].

The volume fraction parameter for the largest nanoparticles can be derived as below:

$$\varphi_V^L = \varphi_V \int_{r_L}^{+\infty} f(r)r^3 \mathrm{d}r \Big/ \int_0^{+\infty} f(r)r^3 \mathrm{d}r \tag{7}$$

where $r_L$ is given by:

$$\int_{r_L}^{+\infty} f(r)\mathrm{d}r = \Phi_L \tag{8}$$

Such size of nanoparticles corresponds to the Brownian relaxation type only.

## 2. Results

### 2.1. Experimental observations

The TEM micrographs at two different scales (Fig. 2) were processed and analyzed with the image processing program ImageJ [38]. The obtained mean nanoparticle diameter was calculated as $\bar{d} = 2 \int_0^{+\infty} f(r) \, r \mathrm{d}r = 11.5$ nm, which comprises a ferromagnetic core diameter $d^m$ and a double nonmagnetic surface layer with a thickness of $a_0 \approx 0.8397$ nm corresponding to the the magnetite cubic spinel structure with the space group Fd3m above the Verwey temperature [39,40]. The mean diameter $\bar{d}$ does not include the oleic acid layer, which is inadequately visible with TEM technique. The oleic acid surfactant layer depth is $\delta \approx 1.97$ nm; hence, the total mean hydrodynamic diameter determined using TEM technique was $\bar{d}_h = 15.44$ nm. The dynamic light scattering experiment provided a larger value of $\bar{d}_h = 19.2$ nm. This discrepancy can be explained by aggregation phenomenon. Therefore, the results based on TEM micrographs can be considered as more reliable in this study. We considered a lognormal diameter probability density function of $f(r)$ with the mean diameter of $\bar{d} = 11.5$ nm.

The dependency of magnetic field pulse magnitude on time is presented in plot of Fig. 3. The characteristic times for pulse rising and falling edge are ∼ 0.4 and ∼ 0.1 s, respectively. The sample of optical transmission signal, which was examined, decreased after the magnetic field impulse started (Fig. 3). The descending *trend* was kept over the period of $\tau_1$, which depends on the field magnitude and the container thickness (Fig. 4). The container thickness affects the transmitted light intensity minimal magnitude of $I_{min} = I_0 - \Delta I = I(\tau_1)$ (Fig. 5).

The reversal of this *trend* is called as a magneto-optical *extinction inversion* [19] (hereinafter: extinction trend inversion or ETI), which is followed by signal increase and saturation over the time of $\tau_2$ before completion of the magnetic field pulse (Fig. 3). The magnetic field switch-off was followed by some transient response and return to the original state (Fig. 3). The transient response can be treated as an additional (residual) ETI.

The switching on of magnetic field pulse led to the formation of FF columnar aggregates, their growth, and fusion ("zippering" or lateral aggregation [22,23]), observed with CMOS camera (Fig. 6). Well developed aggregate structures (starting from the Fig. 6b) is related to an optical microscope resolution and corresponding image contrast quality. The same type of structure (other scale) took place at Fig. 6a, although one is not very recognizable visually due to instrumental constraints. The aggregates were observed concentrated in the central container area. The processing of video snapshots with image processing program ImageJ provided the mean diameter of columnar aggregates as $\bar{d}_A$ dynamics (Fig. 7).

*2.2. Simulation*

The molecular dynamics simulation describes a nanoscopic FF structure evolution starting from a random distribution of nanoparticles and magnetic field pulse rising edge. The field with a magnitude of $H = 0.5$ kOe was sufficient to suppress the closed magnetic flux structures (coils, rings, ring assemblies, tubes, scrolls, and others) [7,12]. Consequently, an equilibrium FF structure was established at the very beginning of simulation, consisting of parallel linear chains of large nanoparticles with mean diameter (Fig. 8(a),(c)):

$$\bar{d}_L = 2 \int_{r_L}^{+\infty} r f(r) \mathrm{d}r = 21.6 \text{ nm} \tag{9}$$

Corresponding dipole-dipole coupling parameter $\Lambda = \mu_0 m^2 / \left[\left(\bar{d}_L + 2\delta\right)^3 kT\right] = 70.82$ confirms aggregation stability. The Langevin argument $\xi(H) = \mu_0 mH/kT$ is evaluated as $\xi(250 \text{ Oe}) = 11.95$ and $\xi(500 \text{ Oe}) = 23.90$ confirming the columnar aggregates structure aligned with an external magnetic field lines. The conditions

$$\lambda \gg \bar{d}_L \tag{10}$$

justifies model of a columnar aggregate as a continuous medium within classical electrodynamics and scattering theory.

The process for a formation and the lateral coalescence of FF columnar aggregates leads to decrease of aggregates quantity (Fig. 8(b),(d)) and their diameter growth (Fig. 7). The latter value of $\bar{d}_A$ was modified here using a correction coefficient of $C$=1.29, which can be related to the depletion forces [41] produced by smaller nanoparticles or the displacive flocculation produced by the oleic acid molecules [42]. However, more plausible origin of this coefficient is related to the presence of smaller nanoparticles than simulated ones ($d > 2r_L$ =20 nm) in real experiment. As it was theoretically shown in [27], solid aggregates include also smaller nanoparticles from the $d$ range 15 nm – 20 nm. This statement is also confirmed by other reports [31]. Presence of such additional nanoparticles leads to correction value 1.29, which can be obtained by integration of the volume fraction over this diameter range and raising to the ⅓ power:

$$C = \left[\int_{15 \text{ nm}/2}^{+\infty} f(r)r^3 \mathrm{d}r \Big/ \int_{r_L}^{+\infty} f(r)r^3 \mathrm{d}r\right]^{1/3} \tag{11}$$

Also, surrounding of the columnar aggregate by even smaller nanoparticles "cloud" is forming a droplet-like structure [43] which leads to an increase of the visible columnar aggregate cross section. However, this area was cut by the image processing tool.

## 3. Discussion

Hence, in the experiments, we directly observed lateral aggregation of FF columnar structures initiated by rising edge of magnetic field pulse for later stages (Fig. 6) and derived from the molecular dynamics simulation (Fig. 8) for initial stages (before the ETI). The above described simulation duration issues restrict possible calculation evolution timeframes. Hence, only a quantitative indirect match of the experiment and the simulation was presented through the connection (extrapolation) of respective branches of $\bar{d}_A(t)$ dependencies (Fig. 7). Changes of the external magnetic field shift experimental and simulation dependencies $\bar{d}_A(t)$ in the same way: higher field increases the aggregate diameter. Regardless of different external field magnitudes (either 250 or 500 Oe), the experimental ETI approximately corresponds to same value:

$$\bar{d}_A(\tau_1) \sim 2\lambda \tag{12}$$

This corresponds to the Mie scattering [16,44] preponderance condition for the cylindrical nanoparticles [45,46] implying model of the columnar aggregate, taking into account the relation (10). A modification of Mie scattering condition here is a factor 2. Such peculiarity could be related to an image digital processing and, also, to not fully rigid structure of visible aggregate cross section outline as a cloud of small nanoparticles surrounded the aggregate; i.e. an effective diameter could be less. Further aggregate growth leads to the optical transmission increase. Here, larger external magnetic field (0.5 kOe) produces a hysteresis nature of the dependence $I(\bar{d}_A)/I_0$, which also could be related to a non-rigid aggregate structure and its corresponding hardening with the lapse of time (Fig. 9).

Variation of FF layers optical transmission due to magnetic field pulse (Fig. 3) can be explained using the following model. The FF phase was observed in suspension state before switching on the magnetic field. The container comprises only separate ferromagnetic nanoparticles (with a mean diameter $\bar{d}$ of 11.5 nm) and small primary aggregates (based on visual observations: $\bar{d}_P < \lambda$) mostly consisting of large nanoparticles (9) ($\bar{d}_L = 21.6$ nm) [7]. As far as the laser wavelength is larger than the size of these inhomogeneities, the magneto-optical extinction corresponds to the Rayleigh branch of the scattering. The field switch-on leads to an increase in the dipole-dipole interaction driven nanoparticles ordering. The columnar FF aggregates start their formation and lateral coalescence leading to the Rayleigh scattering on the cylindrical nanoparticles [45,46]. The aggregates growth during period $\tau_1$ corresponds to the aggregates transverse inhomogeneities size of $\bar{d}_A$; the growth progresses till it reaches the laser beam wavelength order of magnitude. The maximal scattering (minimal value $I/I_0$) corresponding to relation (12) leads to the Mie scattering preponderance, described by Mie-like scattering theory in the cylindrical geometry [45,46]. More precise and comprehensive analysis of such phenomena is provided in Ref. [21,23]. An alternative explanation can be related to the scattering on the nonrigid nanoparticle chains affected by thermal fluctuations [22]. Further aggregates growth and lateral aggregation leads to the relation of $\bar{d}_A \gg \lambda$, corresponding to a geometrical optics spatial scale, consisting of a refraction and a reflection according to Fresnel equations and a minor boundary diffraction. In this case, the extinction reduces significantly, defined almost only by a geometrical shadow effect. Larger aggregates become opaque and observable via camera (Fig. 6). The opaque aggregates shadow effect determines ascending trend for further $I/I_0$ values (Fig. 3). In our opinion, a cross-section (plane XY) of the columnar aggregates was experimentally observed in a perspective projection.

The other effect of first-principle method, which contributes to FF layer optical light scattering, is diffraction on the 2D columnar aggregates grid (Fig. 6) [18]. This grid can be approximately modeled based on a hexagonal close-packing lattice with an edge [47]:

$$a_H = \frac{\pi \bar{d}_A}{3\sqrt{\sqrt{6}\varphi_V^L}} \tag{13}$$

The corresponding dynamics of $a_H(t)$ (Fig. 7) indicates a stronger effect of diffraction mechanism for smaller columnar aggregates in case of an approach to the first (second) ETI point from the left (right) side (Fig. 3). This hexagonal based model expression (13) can be used for estimations of an order of magnitude. However, at least in case of visible in an optical microscope aggregates, it does not correctly reflect the experimental parametric dependency $a_H(H)$ as a function of $\bar{d}_A(H)$ for different values of the magnetic field magnitude $H$. Particularly, the external magnetic field magnitude reduction corresponds to the $\bar{d}_A$ decrease and $a_H$ growth at the same timestamp (c.f. 250 Oe and 500 Oe videos at the given manuscript online version http://dx.doi.org/10.1016/j.jmmm.2016.04.071). This peculiarity can be explained by a columnar aggregates' geometrical dimensions dispersion and their complex fusion with the total sum of nanoparticles redistribution.

The diffraction on the 2D columnar aggregates grid [18] and Mie-like scattering origin [21,23] of the FF extinction have been discussed in other reports. Our simulation confirmed the fact that a spatially homogeneous external magnetic field triggers the formation and lateral coalescence of columnar aggregates. An original result of the present research is a proof that the ETI phenomenon takes place in an uniform external magnetic field. According to Li et al. [20], a spatially inhomogeneous magnetic field is required for an observation of the aggregates fusion and the ETI. The force $\nabla(m_A B)$ was introduced to suppress the columnar aggregates dipole–dipole repulsion, where $m_A$ is a total dipole moment of a single columnar aggregate. However, the present simulation provides the same effect for a homogeneous magnetic field. It can be explained by attraction of linear chains of polarized (magnetized) spheres (known electrostatic or magnetostatic problem [48]), attraction of spontaneous transverse dipole–dipole, and the van der Waals forces between columnar aggregates. Hence, at least at the initial stages of coalescence for the columnar aggregates ($t \leq 0.117$ s), a homogeneous external magnetic field is enough for an ETI phenomenon to take place. This conclusion corresponds to other columnar aggregates observation reports [49] for a field spatial inhomogeneity value $\leq$ 1%.

The observed triggering of a visible aggregation by external magnetic field (Fig. 6) should not be confused with the effects of magnetic field on the liquid–gas (l–g) phase transition [50]. It was shown [27,51] that for a dense phase dissolution [52], the magnetic field reduces the effective coupling between nanoparticles; therefore, an l–g phase transition can occur at a lower temperature of $T_t$. In the present research, the zero magnetic field corresponds to a larger nanoparticles based closed magnetic flux structures. A visual observation is impossible because of small volume fraction of large nanoparticles $\varphi_V^L \sim 0.356\%$. These structures can be considered as small dense phase aggregates. The corresponding stable closed magnetic flux structures have been systematically discussed in the literature [7,12]. The applied external magnetic field in the present research changed the spatial orders from closed magnetic flux aggregates to a columnar structure, which allows an ordered sequential coalescence and a visual observation. This phase transition is unrelated to the *liquid–gas* transition; rather, it is similar to transition from a disordered *liquid* to an ordered *liquid* phase. Both of these phases can be dissolved and transferred to uniform state by increase in temperature. According to our previous observations [27,51], the columnar structure corresponds to a lower phase transition temperature of $T_t$ even though they are visually observable.

There are possible reasons for such columnar structures based on large nanoparticles that have not been experimentally observed in the work [51]: (i) existence of a planar magnetic field geometry, (ii) increased concentration of oleic acid which led to displacive flocculation [42], and (iii) small value of nanoparticle coating rate [27]. The latter two aspects introduces smaller nanoparticles into the aggregates structure. This condition changes the balance between van der Waals and dipole–dipole forces, affecting overall long range phase spatial and magnetic dipole ordering. In the present simulation, the value of $\bar{d}_L = 21.6$ nm corresponds to a dipole–dipole force predominance in separate nanoparticles interactions.

The residual ETI was detected in all investigated samples, and, for examples, was shown for the container of 250 $\mu$m (Fig. 3). It is related to the columnar aggregates destruction by Brownian motion after magnetic field pulse falling edge. The respective mechanism was described earlier; the difference is just a reverse direction of $\bar{d}_A$ change towards trend inversion at the condition (12). However, the asymmetry between optical transmission value at the main ETI and the residual ETI requires special attention. Generally, this asymmetry could be related to:

1.  Conditions of the system are not symmetrical: one case corresponds to the enabled external magnetic field, and other case corresponds to the disabled one. Full temporal symmetry could be in case of a durable linear growth and same durable linear decrease of the external field. However, it was not the case in the given experiment.

2.  Systems with Brownian motion are not reversible systems because they are dissipative. Hence, Rayleigh dissipation function should be considered instead of Lagrange function of the system. Only Lagrange function could produces a reversible process.

In terms of structural analysis, the residual ETI could be analyzed in the following way. After field pulse falling edge, nanoparticles start mutual repulsion due to Brownian motion and a steric layers effect. Columnar aggregates become less dense in contrast to original aggregate growth with continuously supplementing by new nanoparticles packed as a random close-pack media [47]. After the magnetic field pulse falling edge the columnar aggregates promptly transforms into dark regions (Fig. 6). Hence, when such aggregate diameter $d_A$ is close to the wavelength, the aggregate is not a "solid" continuous random close-pack media as one was during main ETI. Numerically, as a result of the aggregate density change, the macroscopic permeability $\varepsilon$ and $\mu$ values are different in case of main and residual ETI. Hence, parameters of Mie scattering are different between main and residual ETI.

The effects of container thicknesses on $\Delta I(H)/I_0$ (Fig. 5) can be explained by a longer extinction optical path inside a thicker FF container. The large field dependence saturation is related to complete destruction of the closed magnetic flux structures within a whole FF bulk. The variations of $\tau_1(H)$, the ETI timestamp at various applied magnetic field magnitudes (Fig. 4), can be due to the faster formation of columnar aggregates and their lateral coalescence inside a thicker FF container. Longer columnar structures, which are possible in the thicker sample, correspond to weaker dipole–dipole repulsion owing to changes of demagnetization tensor [20,53,54]. The strong field saturation of these dependencies relates to the same reason as in the case of $\Delta I(H)/I_0$.

Samples with width 50 $\mu$m and 120 $\mu$m did not show any memory phenomena. However, in case of samples with width 250 $\mu$m, the changes of the optical transmission at different sequential pulse cycles had been observed. Investigation of this phenomenon is of interest for further dedicated research.

## 4. Conclusions

Normal spatially homogeneous magnetic field driven magneto-optical extinction trend inversion in the FF thin layer is caused by transition between the following scattering mechanisms:

- "spherical" Tyndall-Rayleigh scattering in the suspended disordered FF phase: closed-flux aggregates $\bar{d}_P < \lambda$ and separate nanoparticles $\bar{d} \ll \lambda$

- "cylindrical" Rayleigh scattering on the columnar aggregates: $\bar{d}_A \ll \lambda$

- "hexagonal columnar" aggregates lattice based diffraction mechanism $\bar{d}_A < a_H \sim 2\lambda$

- "cylindrical" Mie scattering on the columnar aggregates: $\bar{d}_A \sim 2\lambda$

- a geometrical optics and columnar aggregates shadow: $\bar{d}_A \gg \lambda$

These changes are caused by variation of columnar aggregates diameter $\bar{d}_A$ driven by their growth and lateral coalescence, compared with the laser wavelength. An inhomogeneous magnetic field is not required for this process. The molecular dynamics simulation provided a qualitative match and an asymptotic quantitative match with the experimental observations.

A diffraction mechanism based on hexagonal columnar aggregates lattice also contributes to the magneto-optical extinction trend inversion phenomenon. This contribution predominates in the case of smaller columnar aggregates; hence, it influences the extinction trend inversion time-dependence only on one side of the optical transmission minimum (either the major or the residual one).

Double inversion is possible in FF samples. The residual inversion effect, observed after the magnetic field pulse falling edge, can be explained through a rotational and translational Brownian motion driven columnar aggregates destruction and the corresponding spatially inhomogeneous effective size reduction.

## 5. Acknowledgments

We appreciate Dr. Saeed Doroudiani for useful suggestions and editing. We thank support of NanoMedTech LLC for SEM/TEM for microscopic examinations. We are grateful to Mrs. Daria Tanygina for assistance in processing of images and video.

**Figure captions**

**Fig. 1.** Optical transmission experimental setup: He–Ne laser (#1), electromagnet (#2), FF sample (#3), beam splitter (#4), photodetector (#5), CMOS USB camera (#6) and personal computer (PC).

**Fig. 2.** TEM micrographs at two different scales. The ferrofluid drop was placed on the TEM grid support film of carbon–formvar copper type.

**Fig. 3.** Example of transmitted light intensity and external field magnitude time dependencies for the container thickness of $h = 250$ µm. $I_0$ is the incident light intensity.

**Fig. 4.** Effects of applied magnetic field magnitude on extinction trend inversion timestamp for different container thicknesses.

**Fig. 5.** Variations of transmitted light intensity reduction at extinction trend inversion timestamp as a function of applied magnetic field magnitude for different container thicknesses.

**Fig. 6.** Example of CMOS camera snapshots (plane XY) and real time video clip (in online version) for a prolonged normal direction magnetic field pulse. Timestamps $t$ are pinpointed since the magnetic field $H = 0.5$ kOe pulse rising edge ($t = 0$ s). Container thickness is $h = 250$ µm.

**Fig. 7.** Mean diameter for columnar aggregates $\bar{d}_A$ and their 2D hexagonal lattice edge $a_H$ growth after the magnetic field 0.25 kOe or 0.5 kOe pulse rising edge ($t = 0$ s) in the $h = 250$ µm sample: experimental and simulation results. The $a_H$ is a distance between neighbour aggregates in the plane XY, which is calculated by the equation (13) and is not directly measured in the given experiment (see *Discussion* for further details).

**Fig. 8.** Example of orthographic projections of simulated ferrofluid volume. Timestamps $t$ are pinpointed since the magnetic field $H = 0.5$ kOe pulse rising edge ($t = 0$ s). Field was directed along axis Z. Magnetic moments were approximately aligned with the field direction.

**Fig. 9.** Transmitted light intensity as a function of the columnar aggregate diameter for the container thickness of $h = 250$ μm and different applied magnetic field magnitudes (250 and 500 Oe). $I_0$ is the incident light intensity. The hysteresis corresponds to a continue of aggregate evolution in scope of the external magnetic field pulse timeframes.

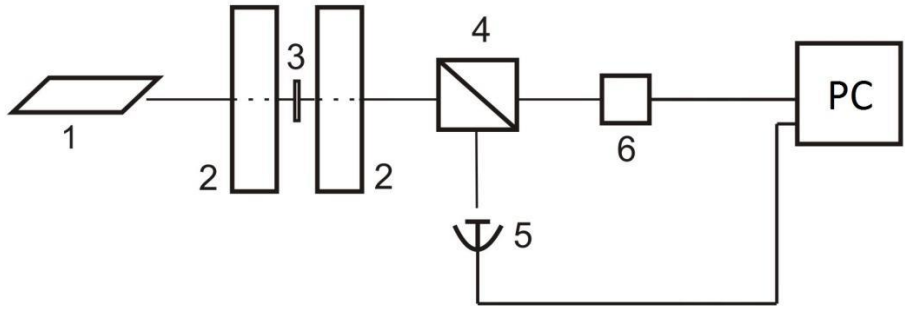

Figure (1)

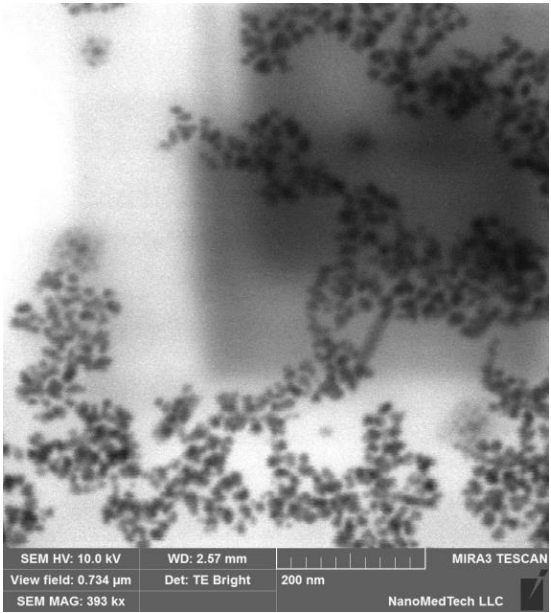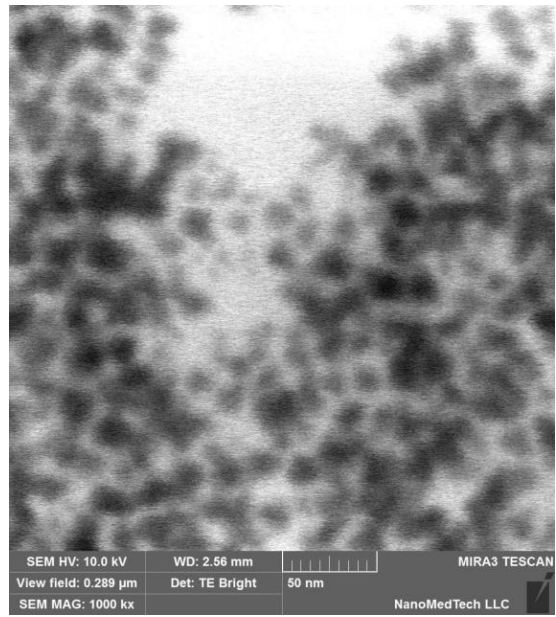

Figure (2)

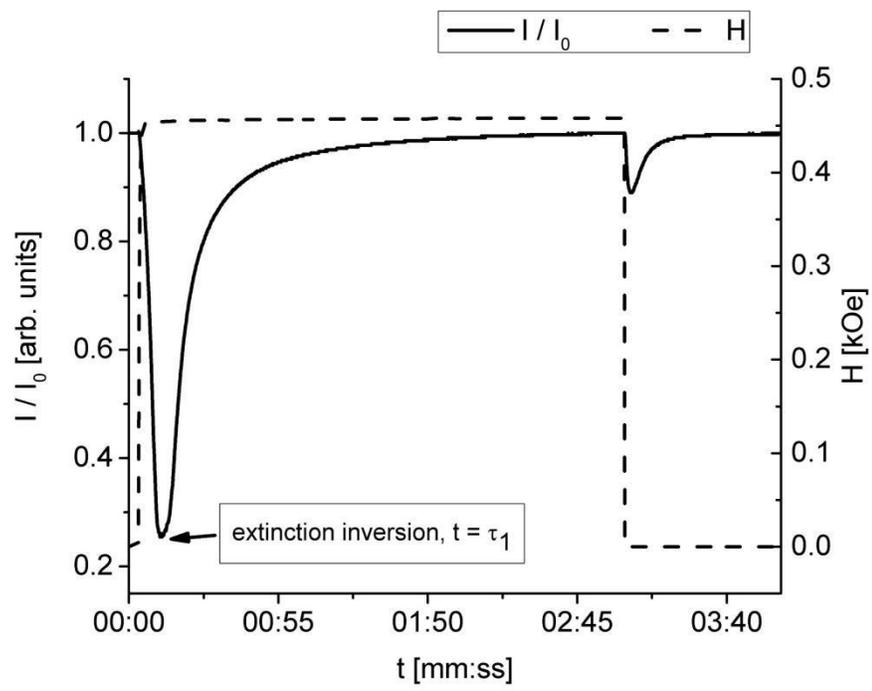

Figure (3)

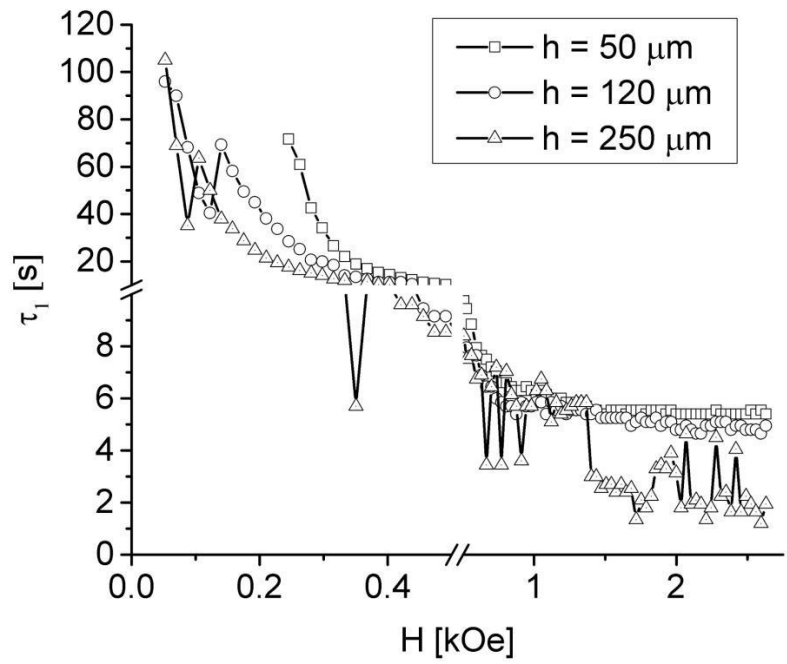

Figure (4)

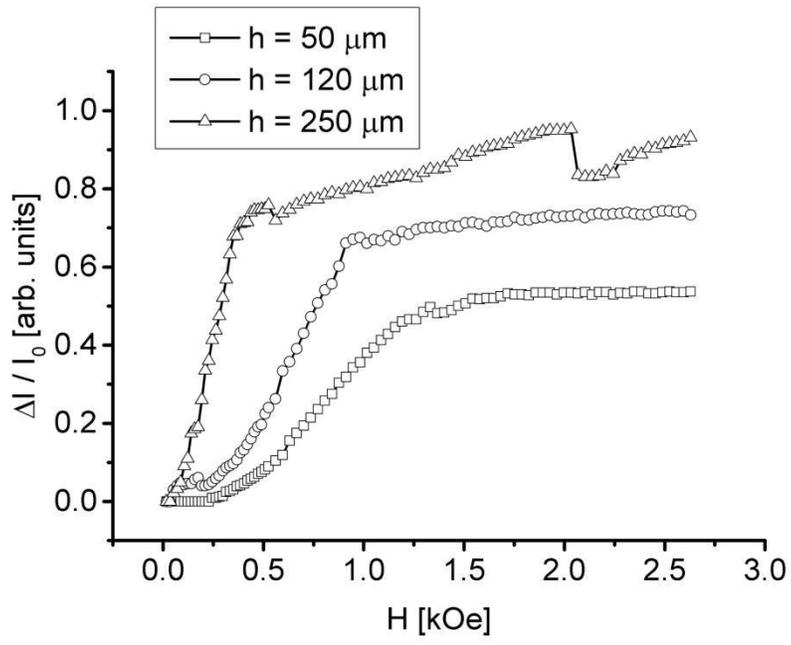

Figure (5)

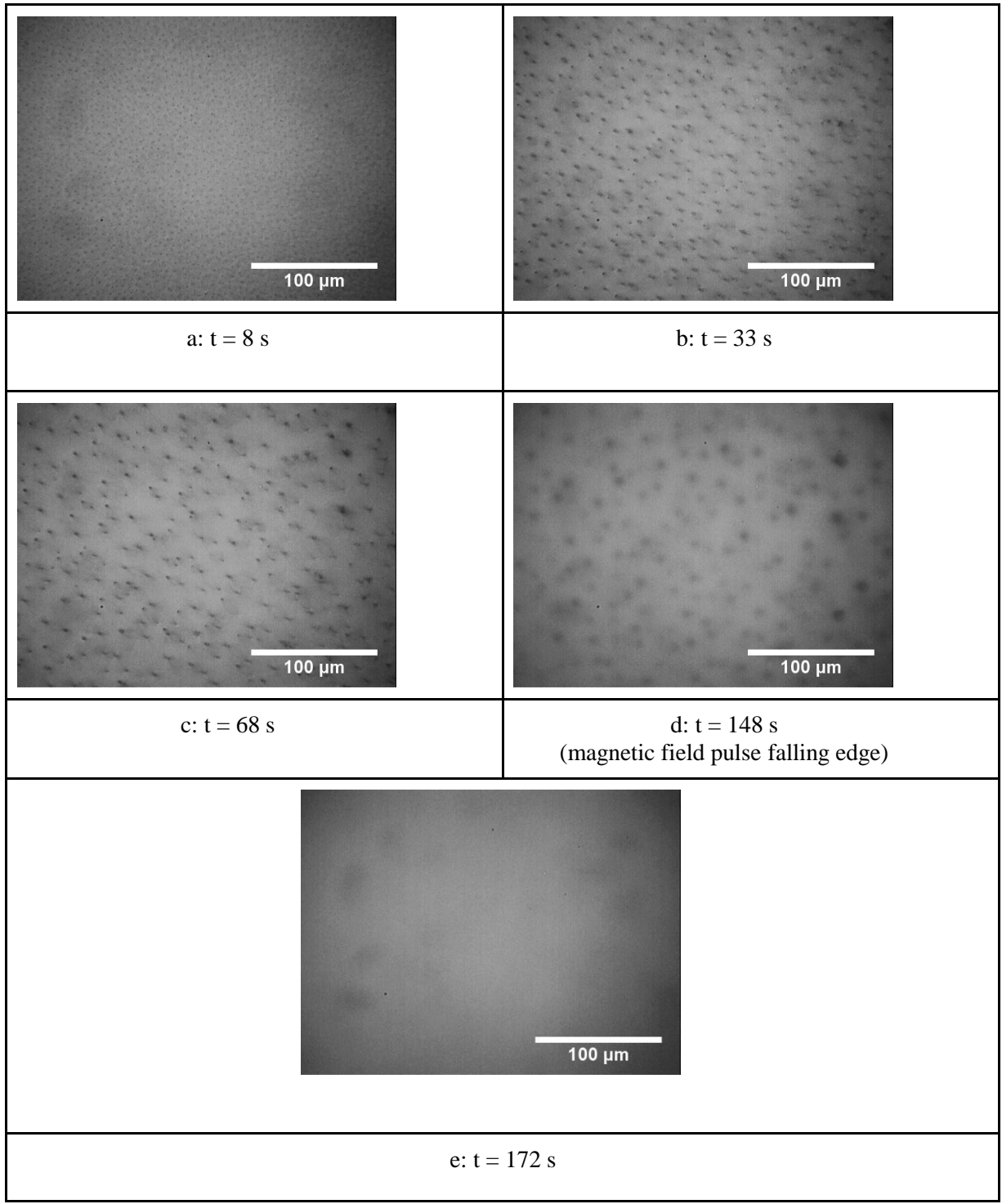

a: t = 8 s

b: t = 33 s

c: t = 68 s

d: t = 148 s
(magnetic field pulse falling edge)

e: t = 172 s

Figure (6)

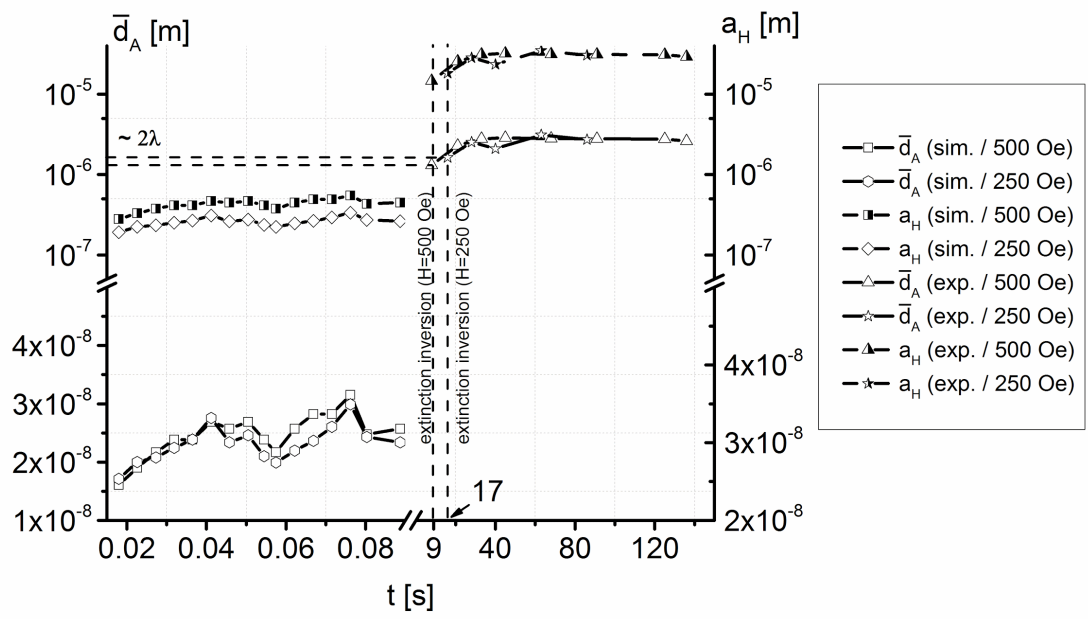

Figure (7)

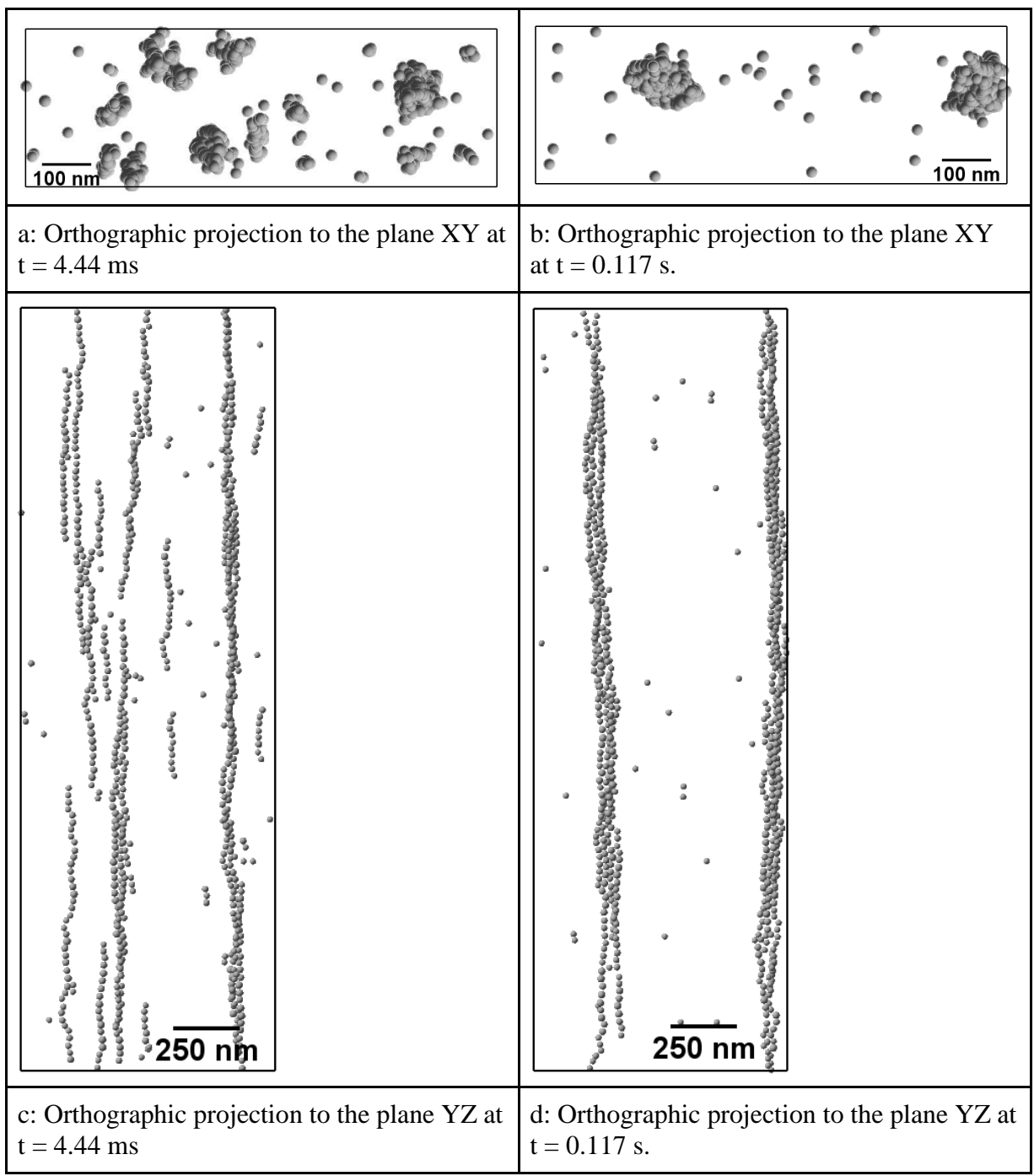

| | |
|---|---|
| a: Orthographic projection to the plane XY at t = 4.44 ms | b: Orthographic projection to the plane XY at t = 0.117 s. |
| c: Orthographic projection to the plane YZ at t = 4.44 ms | d: Orthographic projection to the plane YZ at t = 0.117 s. |

Figure (8)

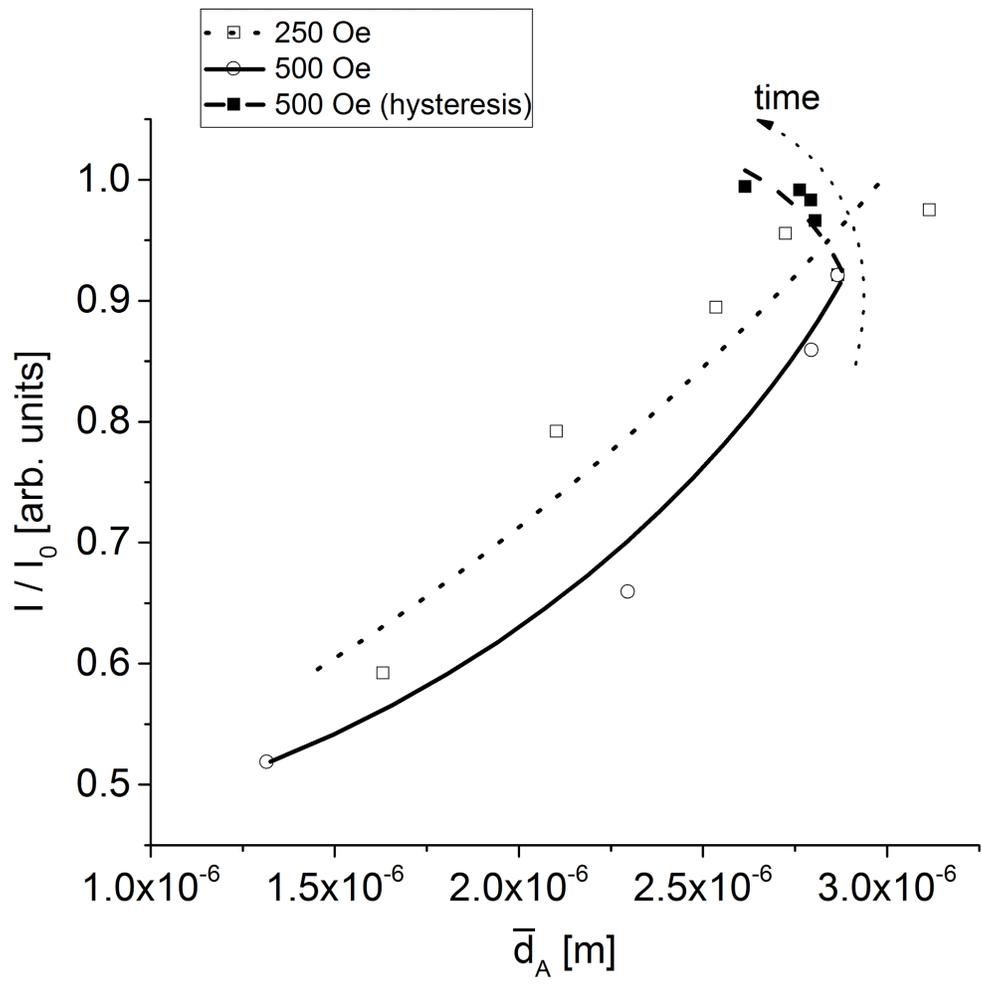

Figure (9)